\documentclass[pra, reprint, superscriptaddress]{revtex4-1}

\usepackage{amsmath,amssymb}	
\usepackage{graphicx}

\begin{document}

\title{Stochastic and deterministic switches in a bistable polariton
  micropillar under short optical pulses}

\author{A.\,V. Uvarov}

\affiliation{Moscow Institute of Physics and Technology, Moscow 117303, Russia}
\affiliation{Skolkovo Institute of Science and Technology, Skolkovo 143025, Russia}
\affiliation{Institute of Solid State Physics, RAS, Chernogolovka 142432, Russia}

\author{S.\,S. Gavrilov}
\author{V.\,D. Kulakovskii}

\affiliation{Institute of Solid State Physics, RAS, Chernogolovka 142432, Russia}
\affiliation{National Research University Higher School of Economics,
  101000 Moscow, Russia}

\author{N.\,A. Gippius}

\affiliation{Skolkovo Institute of Science and Technology, Skolkovo 143025, Russia}
\affiliation{Institute of Solid State Physics, RAS, Chernogolovka 142432, Russia}

\date{\today}
       
\begin{abstract}
 Optical bistability of exciton polaritons in semiconductor
 microcavities is a promising platform for digital optical devices.
 Steady states of coherently driven polaritons can be toggled in tens
 of picoseconds by a short external pulse of appropriate amplitude
 and phase. We have analyzed the switching behavior of polaritons
 depending on the pulse amplitude, phase, and duration. The switches
 are found to change dramatically when the inverse pulse duration
 becomes comparable to the frequency detuning between the driving
 field and polariton resonance. If the detuning is large compared to
 the polariton linewidth, the system becomes extremely sensitive to
 initial conditions and thus can serve as a fast random-number
 generator.
\end{abstract}

\maketitle

\section{Introduction}

This study is concerned with the transient processes accompanying
nonequilibrium transitions in bistable cavity-polariton systems.
Polaritons are short-lived composite bosons that appear due to the
strong coupling of cavity photons and excitons.  They can form
conventional Bose-Einstein condensates fed by the excitonic reservoir
under nonresonant excitation conditions~\cite{Deng2010}.  On the other
hand, under direct resonant driving, they form highly nonequilibrium
coherent states ~\cite{Elesin1973sov}.

When the pump frequency \emph{slightly} exceeds the polariton
resonance frequency, the system has two or more steady
states~\cite{Baas2004,Paraiso2010,Amo2010,Gavrilov2010}. Their exact
number depends on pump polarization because the interaction between
polaritons is spin-sensitive: polaritons with parallel spins strongly
repel each other. Varying excitation involves transitions between
steady states, and a number of other switching mechanisms were
proposed as well, based on short optical or acoustic pulses acting as
switch triggers~\cite{Gavrilov2012,Klaas2017,Zhang2015,Cerna2013}. In
the general case, the switch has the shape of a sharp jump followed by
decaying transient oscillations. When their amplitude is small, their
spectrum resembles above-condensate excitations in stationary Bose
condensates \cite{Brichkin2015,gavrilov2015blowup,Solnyshkov2008}. In
particular, when the blueshift of the pumped mode (condensate) equals
the pump detuning and the lifetime is large, then the above-condensate
modes have the Bogolyubov (sound-like) spectrum.

Here we study transient processes with particularly great intensities
of nonlinear oscillations. The condensate is created by a continuous
plane wave and disturbed by short-term optical pulses; for simplicity,
we consider a circularly polarized system with only two steady states
(``ON'' and ``OFF'' for brevity) in a finite range of cw pump powers.
A high-intensity switching pulse involves oscillations between these
steady states and it becomes difficult to predict which of the two
will be an eventual solution after the oscillations have decayed.  We
analyze this depending on the pulse parameters, its phase and
intensity.

Two different types of solutions exist.  First, the pulse intensities
resulting in the ON and OFF states can be tightly interlaced in the
phase space.  The intensity variation which is sufficient for altering
the eventual steady state becomes infinitesimal as the polariton decay
rate tends to zero. Thus, for a pulsed laser with a finite accuracy of
repetition, this regime effectively enables a random number
generation.  The relative probabilities of the two outcomes are found
to depend on the cw pump intensity.

The second type of solutions is, by contrast, completely
deterministic.  Within another area of the phase space, the system is
guaranteed to be in the OFF state after the pulse has gone.  The two
types of solutions can coexist for the same set of the microcavity
parameters. Therefore, the pulses with a ``deterministic'' outcome can
be used to reinitialize the proposed random number
generator. With increasing pulse power, stochastic and
deterministic regions are well separated in the phase space but still
alternate one after another.


\section{The model} \label{sec:setup}

\subsection{Gross-Pitaevskii equation}

Polaritons in semiconductor microcavities are formed due to the strong
interaction of cavity photons and two-dimensional excitons confined in
quantum wells.  This system is excited by a plane wave with frequency
$\omega_p$.  Polariton spectrum is split into two dispersion curves,
lower and upper polariton branches:
\begin{align}
\label{eq:dispersion}
\omega_{LP,UP} = & \frac12 \left[ \omega_{cav}(\mathbf{k})+\omega_{exc}(\mathbf{k}) \right] \nonumber \\ 
                 & \mp \sqrt{(\omega_{cav}(\mathbf{k})-\omega_{exc}(\mathbf{k}))^2 + \Omega_R^2},
\end{align}
where $\Omega_R$ is the Rabi splitting. 

In order to localize polaritons at a certain place, the cavity mirrors
can be etched out except the micropillar of a particular size.  This
approach allows one to confine polaritons while retaining a high
$Q$-factor \cite{Schneider2016}. When the pillar has the radius of a
few microns, the continuous spectrum is changed into a set of discrete
energy levels due to size quantization. In planar cavities, the
behavior of the macroscopic wave function $\psi(\mathbf{r},t)$
describing the polariton condensate obeys the Gross-Pitaevskii
equation.  The pillar can be represented by a deep potential well
$U(r)$:
\begin{align}
\label{eq:GPE}
i \frac{\partial \psi}{\partial t} = & (\omega_{LP}(-i\nabla) - i \gamma)\psi + U(r) \psi  \nonumber \\
& + V_a |\psi|^2 \psi 
+ f_a(\mathbf{r},t)e^{-i \omega_p t},
\end{align}
where $V_a$ is the polariton repulsion energy per unit area, $f_a$ is
the pumping amplitude, $\gamma$ is polariton decay rate
(damping). The units of $\psi$ are chosen so that
$\left[ V_a \right]=1$. Here we assume that the pump is circularly
polarized and neglect the spin degree of freedom.

This form of the equation is a simplified model, as we dropped the
upper polariton branch and neglected the dependence of $V_a$ on the
wavevectors of the interacting waves. However, we believe that this
simplification does not invalidate our conclusions because (i) the
pumping frequency considered in this work is close to the lowest
eigenstate, thus the upper polariton branch is energetically too far
to have any significant influence, and (ii) the lowest eigenmode of
the pillar contains the most of the wavefunction density, and it lies
in the $k$-space area where the discrepancy caused by simplification
of $V_a$ is minimal.  Hence, the error appears mostly in the other
modes, which should not affect the result qualitatively. Nonetheless,
one should keep in mind that such simplification effectively enhances
the polariton-polariton interaction, and therefore it only gives the
upper estimate of the impact of remote wavevectors.

The polariton system is known to exhibit bistable behavior if the
pumping amplitude $f$ is constant and the frequency detuning
$D = \omega_p - \omega_{LP}$ is larger than $\sqrt{3} \gamma$. The
switching between such states is realized by adding a short pulse to
the constant driving field:
\begin{equation}
\label{eq:pump_shape}
f_a(\mathbf{r},t) = \left( f_0 + f_1 \cdot 2^{-\frac{(t - t_0)^2}{2 \tau^2}}\right).
\end{equation}
Here $f_1$ is complex-valued, so that it contains information about
phase difference of cw pumping and the pulse:
$f_1 = |f_1| \exp(-i \phi)$. The full width at half-magnitude (FWHM)
of the pulse equals $2 \tau$.

\subsection{Single-mode approximation of a micropillar}



The core assumption underlying the single-mode approximation is that
all other modes are sufficiently far from the pump frequency so that they
do not affect the system behavior. Keeping that in mind and
assuming that pumping is monochromatic (i.e. the time dependence of
$f$ is much slower than the oscillations at the pump frequency), we
introduce an ansatz:
\begin{equation}
    \label{eq:ansatz}
    \psi(\mathbf{r},t) = \Psi(t) u_0(\mathbf{r}) e^{-i \omega_p t} + \chi(\mathbf{r}, t),
\end{equation}
where $u_0$ is the normalized wavefunction of the ground state in a
circular potential well, $\chi$ denotes the deviation of the solution
shape from that of the ground state and is assumed to be small
compared to the first term. Such an ansatz corresponds to exciting the
system with a very weak pump, so that the nonlinear term can be
neglected. Substituting (\ref{eq:ansatz}) into (\ref{eq:GPE}) and
projecting it on $u_0$ yields a single-mode equation for $\Psi$:
\begin{equation}
    \label{eq:single-mode_eff}
    i \frac{\partial \Psi}{\partial t} =
    (-D - i\gamma)\Psi
    + V |\Psi|^2 \Psi
    + f,
\end{equation}
where
$V = V_a \int \! |u_0|^4 \, \mathrm{d}S , \quad f = \int \! u_0^* f_a
\, \mathrm{d} S$.


\section{Single-mode approximation} \label{sec:single-mode}

\subsection{Reaction to the pulse}

\begin{figure}
    \includegraphics[width=\linewidth]{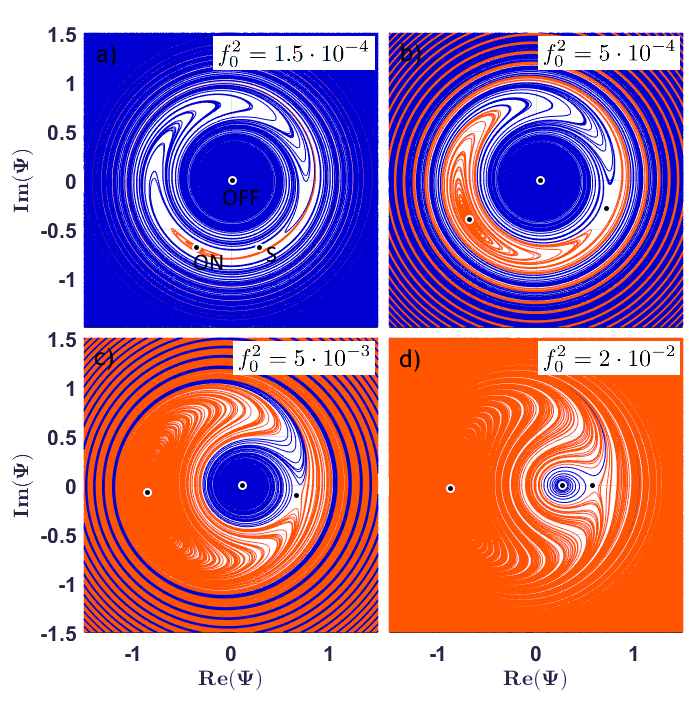}
    \caption{Phase plane for equation (\ref{eq:single-mode_eff}) with
      time-independent pumping amplitude. Trajectories shown by red
      and blue curves are attracted to the ON and OFF states,
      respectively. Points highlight the stationary solutions: ON,
      OFF, and $S$. Here and in all other figures, unless stated
      otherwise, $\hbar D = 0.6$ meV, $\hbar \gamma = 0.014$ meV,
      $V = 1$.}
    \label{fig:Phase_plane}
\end{figure}

In the absence short-term pulses, equation (\ref{eq:single-mode_eff})
is autonomous. Its typical phase portrait is depicted in
Fig. \ref{fig:Phase_plane}. The two steady states are stable foci
(labeled as ON and OFF), while the unstable state is a saddle point
(S).  The $\Psi$ phase plane is shown by the trajectories according to
their final state and shows the basins of attraction of two steady
states.

Now let us consider the system excited with a short pulse which then
relaxes to one of two steady states. Naturally, the resulting state
depends on the properties of the pulse.  This problem has three
characteristic time scales: the decay time $\gamma^{-1}$, the pulse
duration $\tau$, and the inverse frequency of rotation in the phase
space, $T = |-D +V|\bar\Psi|^2 \bar\Psi|^{-1}$, where $\bar\Psi$
denotes a characteristic value of $\Psi$, e.g. its average absolute
value during the pulse.
%
The solution changes the most noticeably upon varying the ratio
between the last pair of characteristic times, whereas $\gamma^{-1}$
is assumed to largely exceed both of them. If $\tau \ll T$, the
solution does not propagate too far in the transverse direction during
the pulse. With sufficient accuracy, the pulse can be replaced by the
delta function. 
Since the frequency depends on $|\Psi|^2$, this
approximation is valid only in a finite interval of amplitudes.  Its
validity is also limited by the uncertainty relations: if the pulse is
short, its spectral width is high, thus the pulse will necessarily
affect the eigenstates beyond the first one.

\begin{figure}
        \includegraphics[width=\linewidth]{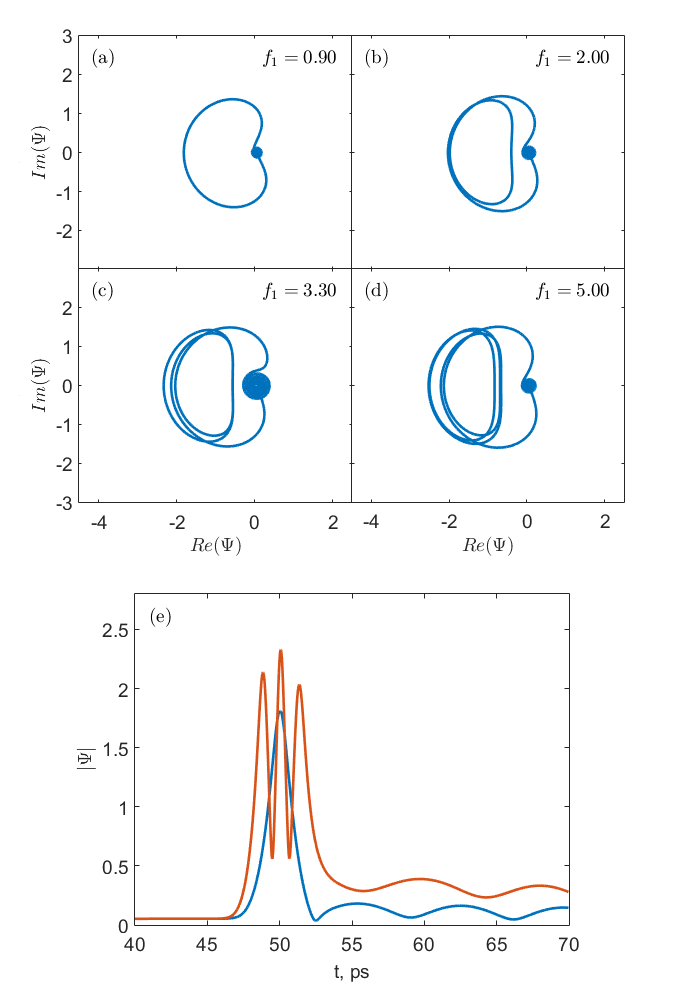}
        \caption{(a) - (d): Four different trajectories that leave the
          system in the OFF state. (e): trajectories for $f_1=0.9$
          (blue) and $f_1 = 3.3$ (orange) versus time. Despite making
          several loops with high $|\Psi|$ during the switching pulse,
          all these trajectories converge to the OFF state. Note as
          well that the central point of such rotations changes slowly
          compared to the pulse amplitude $f_1$. The FWHM of the
          pulses is 2 ps.}
    \label{fig:four_belts}
\end{figure}

On the contrary, if $\tau \gg T$, then the driving field and phase plane change slowly,
while $\Psi$ evolves relatively fast.
 If, in addition, $\tau$ is comparable with $\gamma^{-1}$, the
solution follows the equilibrium state adiabatically. In this case, the
final state depends on how the system leaves the bistability region
during the pulse. If the pump power is beyond the bistability turning
point, the OFF state merely vanishes, and thus the system can be
reliably delivered to the ON state. The only exception occurs when the
phase difference between the background cw pump and the pulse
approximately amounts to $\pi$.  Here the two pump sources cancel each
other and the overall intensity drops down to quite a low value before
it restores back to the cw intensity.  As a result, the system is
reliably delivered to the OFF state. This regime is consistent with
switching discussed in Ref.~\cite{Zhang2015}.

Now let us consider the regime when $\tau \sim T$. During the
switching pulse, one can imagine an instant phase plane with instant
steady states and their basins of attraction. The absolute value of
the ON point moves rapidly at the leading and back fronts of the pulse
and slows down near the pulse maximum. Due to the strong blue shift of
the resonance at high powers, $\Psi_{ON}$ grows as a cubic root of the
pumping amplitude. As a result, during the switching pulse, the
solution rotates around the slowly drifting ON state corresponding to
the instantaneous magnitude of the total external field. After the
switching pulse has turned off, $\Psi$ gradually returns to one of the
stationary states following the flow lines of the stationary phase
plane.

This switching regime can be illustrated by the example of a
rectangular switching pulse $f_1$ turned on during a certain time
$\tau$. The solution rotates around $\Psi_{ON}$ with negligible
decay, and after the pulse has turned off, the solution goes along the
stationary phase trajectories. If for two pairs $(f_1, \tau)$ the
angle spanned by the solution around the ON point is the same up to
$2 \pi$ (see an example of such trajectories on
fig. \ref{fig:four_belts}), the resulting trajectories are the same
and these pairs can be considered equivalent.
The angular velocity of such rotation can be estimated by considering a
perturbed solution near the ON state: $\Psi = \Psi_{ON} + \delta
\Psi$. If we substitute this into (\ref{eq:single-mode_eff}) and keep
only the terms no smaller than $\delta \Psi$, we get
\begin{equation}
i \dot{(\delta \Psi)} = (-D - i \gamma) \delta \Psi
 + V \left( |\Psi_{ON}|^2 \delta \Psi + \Psi_{ON}^2 \delta \Psi^{*} \right).
\end{equation}
Let $\delta \Psi = r e^{i \theta}$; then, after separating the real
and imaginary parts of the equation, we obtain
\begin{align}
\dot{r} &= V \mathrm{Im}(\Psi_{ON}^2 e^{-2 i \theta})r - \gamma r; \\
\dot{\theta} &= (D - 2 V |\Psi_{ON}|^2) + V \mathrm{Re}(\Psi_{ON}^2 e^{-2 i \theta}).
\end{align}
The latter equation suggests that in the case of a very strong
pumping, the solution rotates clockwise (that is, $\dot{\theta} < 0$),
and $-3V|\Psi_{ON}|^2 < \dot{\theta} < -V|\Psi_{ON}|^2$. Taking into
account that, asymptotically, $|\Psi_{ON}|^2 \sim f_1^{2/3}$, we have 
$|\dot{\theta}| \sim |f_1|^{2/3}$.  If the solution has made a full
circle around the $\Psi_{ON}$ point, i.e.
$2 \pi n = \tau \dot{\theta} \sim \tau f_1^{2/3}$, the corresponding
trajectories are equivalent. Thus, to estimate the amplitudes of the
equivalent switching pulses in the phase space, we should solve this
equation for $f_1$, which eventually gives $f_1^{(n)} \sim n^{\frac{3}{2}}$.

\subsection{Simulation results}

\begin{figure}
    \includegraphics[width=\linewidth]{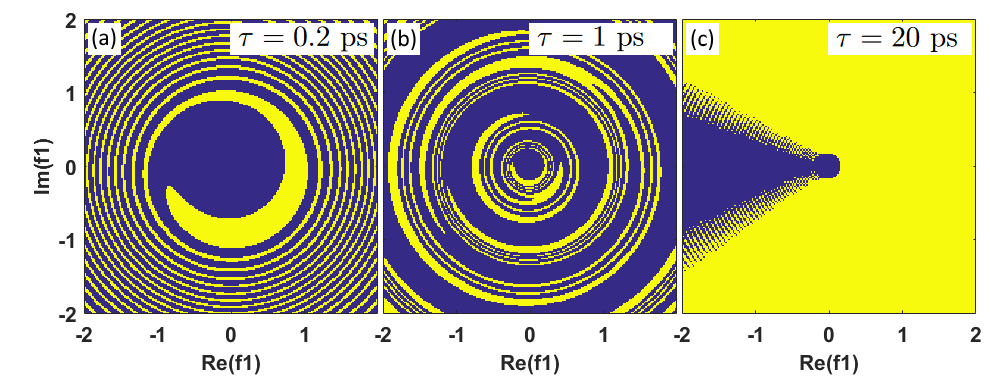}
    \caption{Final state of the single-mode system as a function of
      complex amplitude of the pulse. Starting state is OFF. Different
      panels show the result for different $\tau$.
      $f_0^2 = 5 \cdot 10^{-4}$. All other parameters are fixed and
      equal to those in fig. \ref{fig:Phase_plane}.}
    \label{fig:single-mode_limits}    
\end{figure}

\begin{figure}
        \includegraphics[width=\linewidth]{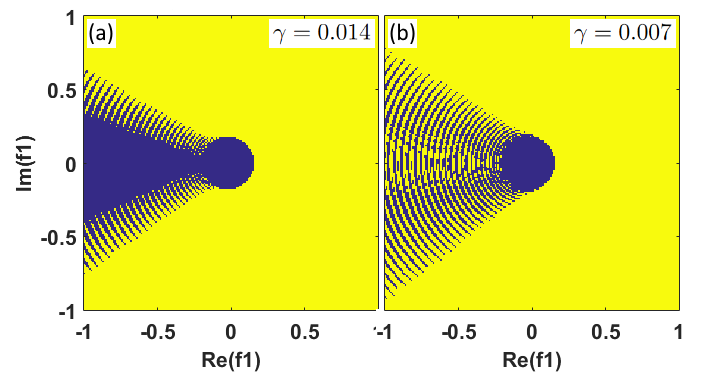}
        \caption{Serrate edge of the regions in the $f_1$ plane
          changes with different decay ratio.}
    \label{fig:long_pulse_gammas}    
\end{figure}

We simulated the response of the single-mode system to the pulses
discussed above. Each set of the initial conditions and other system
parameter results in one of two states, ON and OFF, which can be
indicated by different colors on the phase plane.  The result of such
mapping can be seen in Fig. \ref{fig:single-mode_limits}. In the
series of subplots, the pulse duration increases from left to
right. The leftmost panel shows the final states in the case of very
short pulses. Such a short pulse makes the system evolve along the
straight line (that is why the look of the $f$ plane resembles the
$\Psi$ plane, drawn in accordance with the basins of attraction). The
rightmost panel shows the other extreme condition. Here the different
sectors of the plane almost do not mix with each other, because the
solution follows the ``equilibrium'' adiabatically. When $f_1$ is
aligned in phase with the cw pumping, the OFF state becomes
unavailable, and so the solution sticks with the ON state. When $f_1$
is aligned against the cw pumping, there is a time interval when the
ON state is unavailable, so the system returns to the OFF state. The
fact that the system is still non-adiabatic manifests itself in the
serrate border of the regions: the transient oscillations do not
smooth out completely during the pulse. The longer the lifetime, the
more pronounced the serrate area of the phase space (see
Fig. \ref{fig:long_pulse_gammas}).

\begin{figure}
    \includegraphics[width=\linewidth]{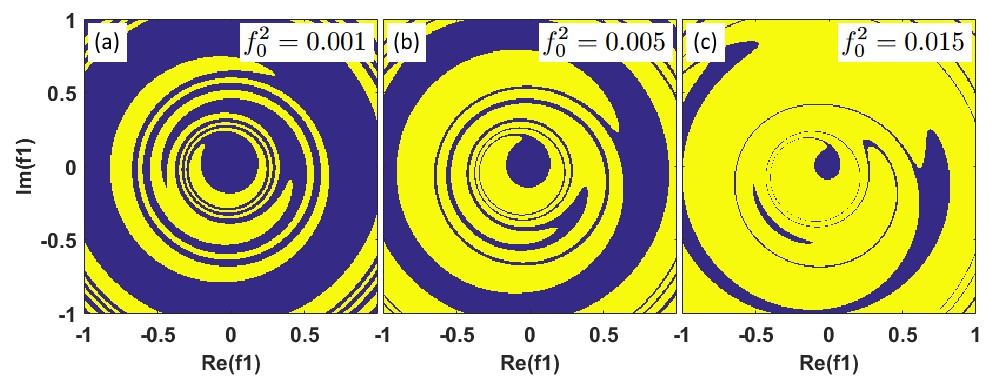}
    \caption{Evolution of the middle panel in
      fig. \ref{fig:single-mode_limits} under gradual increase of the
      background cw  pumping.}
    \label{fig:ring_vs_power}
\end{figure}

The middle panel shows the final states for the intermediate pulse
length. This picture exhibits a small region in the middle where the
characteristic spiral unwinds counterclockwise and then back
again. This behavior is consistent with the explanation suggested
above. With increasing $|f_1|$, the trajectories firstly get to higher
values of $|\Psi|$. Then the growth stops (as the rotation
establishes) and after the pulse the value of $|\Psi|$
decreases. Finally, at certain $|f_1|$, the system arrives in the
vicinity of the OFF state, regardless of the relative phase
$\phi$. That is the reason for the dark ring in seen in the
Figure. This pronounced ring-shaped area belongs to the OFF basin, and
as long as the trajectory ends up close to zero, the result does not
depend on the relative phase of the pulse. However, with the
increasing the background cw pumping the OFF basin shrinks and moves
off-center, which leads to the gradual disappearance of the ring (see
fig. \ref{fig:ring_vs_power}).  It is important to note that the
coloring repeats itself indefinitely, albeit with some transverse
distortion. Again, this is predicted by the model of the solution
revolving around the instant equilibrium. 

\begin{figure}
\includegraphics[width=\linewidth]{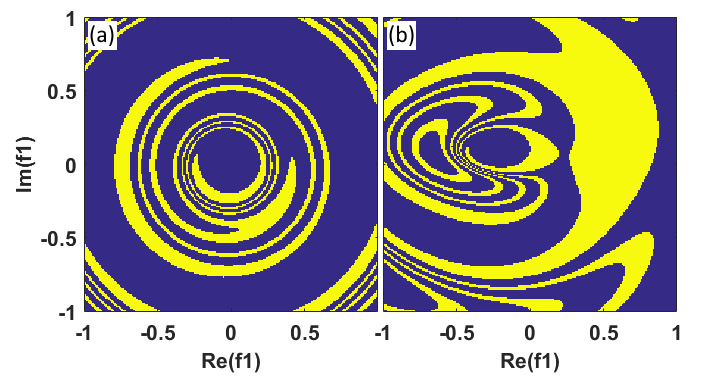}
\caption{Final states of the system (after the pulse) in the cases
  when the starting state is OFF (a) and ON (b).
  $f_0^2 = 0.0005\ \mathrm{meV} \cdot \mathrm{ps}^{-2}$, pulse FWHM =
  2 ps.}
\label{fig:spirals}
\end{figure}

The smaller the decay ratio, the more rotations the solution makes
before arriving into the saddle point and ``scattering'' into the
vicinity of a certain focal point. In the sense of
Fig. \ref{fig:spirals}, this means more ``coils'' in the diagram. The
rest of the diagram stays more or less stable, including the ends of
the spirals, the ring with the guaranteed OFF state, etc.

\begin{figure}
    \includegraphics[width=\linewidth]{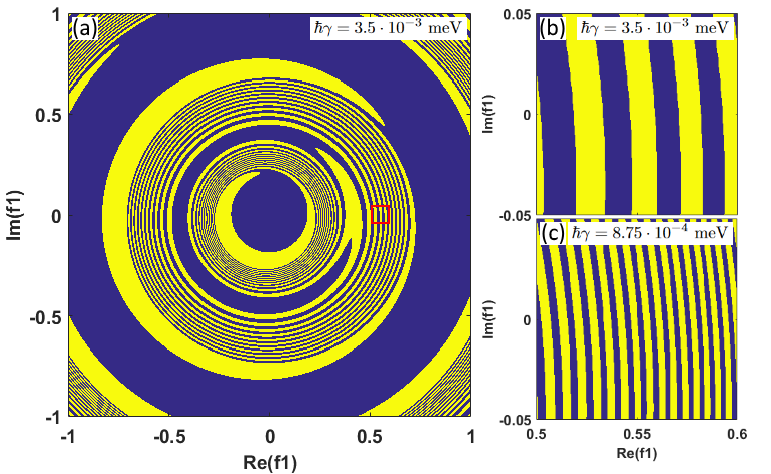}
    \caption{(a): Final states diagram for a value of $\gamma$ much
      smaller than other characteristic times of the system. Panels
      (b), (c) magnify the area highlighted with the red square, for
      different values of the decay parameter. $\tau = 1 \text{ps}$,
      $f_0^2 = 5 \cdot 10^{-4}$.}
    \label{fig:rng_pic_and_lines}    
\end{figure}

In the limit of $\gamma \rightarrow 0^{+}$, there is a certain region
where the density of the ``coils'' tends to infinity
(fig. \ref{fig:rng_pic_and_lines}). In this regime, the system
effectively becomes a random number generator unless the pump power is
known with an infinite accuracy. In general, this generator is biased
to one of the two states depending on the background pump parameters. This
opens a way to manipulate the relative probability of the
outcomes. Surprisingly, in the phase space, this regime coexists with
the completely deterministic regime.

The approach developed above can be
applied to the single-mode system in the ON state as well. When the
system is in the ON state, the pulse can drive it down or leave in the
ON state, depending on the pulse parameters. However, since the ON
state is substantially off-center in the phase diagram
(Fig. \ref{fig:Phase_plane}), there is no analog of the ``belt'' that
is seen when the system is initially in the OFF state (see
fig. \ref{fig:spirals}, (b)). Thus, if we were to send the system to
the OFF state, we could apply the pulse with the amplitude from that
belt.  Such pulses can drive the system down or leave unchanged, but
they definitely cannot drive it up.  Thus, if we cannot control the
phases of the pulses, we still can send a train of pulses and
guarantee that the system is sent down after a certain number of them.

\section{Conclusions and outlook} \label{sec:conclude}

We have studied the driven polariton mode under the joint action of cw
and pulsed pumping when the system is bistable and the short pulse
acts as a trigger of transitions between steady states. The result is
shown to be a nontrivial function of the pulse duration and amplitude.
Depending on these parameters, several dynamical scenarios can
be distinguished.

First, for sufficiently small decay rate, there is a region of pumping
amplitudes where any point is within a small distance from both basins
of attraction. Second, there is a region of large pumping amplitudes
where the transition from the OFF state to the ON state is forbidden
for any pulse phase. The combination of these two effects can be used
to make a random-number generator: first, in response to the short
pulse the system goes to one of two states that cannot be predicted in
view of finite accuracy, and, second, the system can be reinitialized
by the pulses whose amplitudes are guaranteed to not switch the system
into the ON state. However, such a device would require a long
lifetime of polaritons and correspondingly slow rate of transitions.

At the first glance, the polariton oscillator behaves similarly to a two-level
quantum system under pulsed excitation which experiences the Rabi oscillations. 
However, the two-level system periodically passes through 
both excited and ground states, while the discussed
polariton oscillator orbits the ON state and does not pass through the
low-intensity OFF state during the pulsed excitation.

\section{Acknowledgements}

The work was supported by the Russian Science Foundation (Russian Federation) (Grant No. 14-12-01372).
The images in this work were prepared using \textsc{Matlab} \cite{MATLAB:2016}.


%

\end{document}